\documentclass[%
 reprint,
 amsmath,amssymb,
 aps,
prb,
 longbibliography,
 floatfix,
]{revtex4-1}

\usepackage{graphicx}
\usepackage{dcolumn}
\usepackage{bm}
\usepackage{hyperref}

\begin{document}


\title{'Checkerboard stripe' electronic state on cleaved surface of NdO$_{0.7}$F$_{0.3}$BiS$_{2}$ probed by scanning tunneling microscopy}

\author{T. Machida$^{1}$, Y. Fujisawa$^{1}$, M. Nagao$^{2,3}$, S. Demura$^{3}$, K. Deguchi$^{3}$, Y. Takano$^{3}$ and H. Sakata$^{1}$}

\affiliation{$^{1}$Department of Physics, Tokyo University of Science, 1-3 Kagurazaka, Shinjuku-ku, Tokyo 162-8601, Japan\\
$^{2}$1University of Yamanashi, 7 -32 Miyamae, Kofu 400-8511, Japan\\
$^{3}$National Institute for Materials Science, 1-2-1 Sengen, Tsukuba, Ibaraki 305-0047, Japan}

\begin{abstract}
We present scanning tunneling microscopy measurements on a cleaved surface of the recently discovered superconductor NdO$_{0.7}$F$_{0.3}$BiS$_{2}$ with a transition temperature ($T_{\mathrm{c}}$) of 5.1 K.
Tunneling spectra at 4.2 K (below $T_{\mathrm{c}}$) and 22 K (above $T_{\mathrm{c}}$) show a large spectroscopic gap ($\sim$40 mV), which is inconsistent with the metallic nature demonstrated in bulk measurements. Moreover, we find two interesting real-space electronic features. The first feature is a `checkerboard stripe' electronic state characterized by an alternating arrangement of two types of nanocluster. In one cluster, one-dimensional electronic stripes run along one Bi-Bi direction, whereas, in the other cluster, the stripes run along the other Bi-Bi direction. The second feature is a nanoscale electronic inhomogeneity whose microscopic source seems to be atomic defects on the cleaved surface or dopant F atoms.
\end{abstract}

\maketitle
In recent decades, several layered compounds have generated much sustained interest, because of their peculiar phenomena. An example of these phenomena is charge-density-wave (CDW) formation in the transition metal dichalcogenides\cite{DEMoncton,MNaito,ASoumyanarayanan,BSipos,LJLi,JJKim,RAng} and trichalcogenides\cite{NRu,HJKim}, in which the superconductivity (SC) often coexists with the CDW\cite{DEMoncton,MNaito,ASoumyanarayanan,BSipos,LJLi,JJKim,RAng}.
Because of the coexistence of CDW and SC, these materials are important for understanding the interplay between these phenomena\cite{BSipos,LJLi,RAng}.
Another example is the high-temperature superconductivity in copper oxides\cite{Bednorz} and iron-pnictides\cite{Kamihara}. The superconductivity in these materials often coexists with or emerges in the proximity of some electronic ordered states\cite{SAKivelson,CVParker,MJLawler,AMesaros,YKohsaka,TMChuang,SKasahara}, which break the $C_{4}$-symmetry of the underlying crystal lattice. Explorations of the relation between SC and the electronic ordered states have offered interesting information for understanding of the mechanisms of unconventional superconductivity.

The recent discovery of BiS$_2$-based superconductors has triggered many research activities, because of (i) the structural similarity to cuprates and pnictides and (ii) their relatively high $T_{\mathrm{c}}$\cite{Mizuguchi_1,SK_Singh,Mizuguchi_2,JXing,RJha_1,Demura,MNagao,XLin,DYazici}.
Recent theoretical calculations have suggested that a quasi-one-dimensional character of the band dispersion provides a good Fermi surface nesting\cite{Usui,XWan,TYildirim,BLi,GBMartins}.
Some of them have proposed that an unstable phonon at $\boldmath{q}$ = (q, q, 0) corresponding to the Fermi surface nesting vector causes a CDW instability\cite{XWan,TYildirim}. How the SC is related to this Fermi surface nesting has been argued vigorously\cite{Usui,XWan,TYildirim,BLi,GBMartins,YGao,YYang,JLee}. Meanwhile, the symmetry of the superconducting gap has also been investigated theoretically and experimentally. These studies have suggested the isotropic $s$-, anisotropic $s$-, and $d$-wave symmetries as possible pairing symmetries\cite{Shruti,GLamura,YGao}. Despite these studies described above, there is less conclusive empirical evidence for determining the pairing symmetry and for uncovering whether CDW instability occurs.

One of the powerful techniques for addressing directly these issues is a scanning tunneling spectroscopy (STS) using a scanning tunneling microscopy (STM), which has been used to determine the pairing symmetries in a large variety of superconductors\cite{Rev_Fischer,Rev_Hoffman,KMcElroy,Hanaguri,MPAllan}.
Additionally, this technique has made a significant contribution not only to visualize the CDW \cite{ASoumyanarayanan,JJKim} and other electronic ordered states\cite{CVParker,MJLawler,YKohsaka,TMChuang} but also to disentangle the relationship between the superconductivity and the electronic ordered states.

\begin{figure*}[tbh]
\begin{center}
\includegraphics[width=12cm]{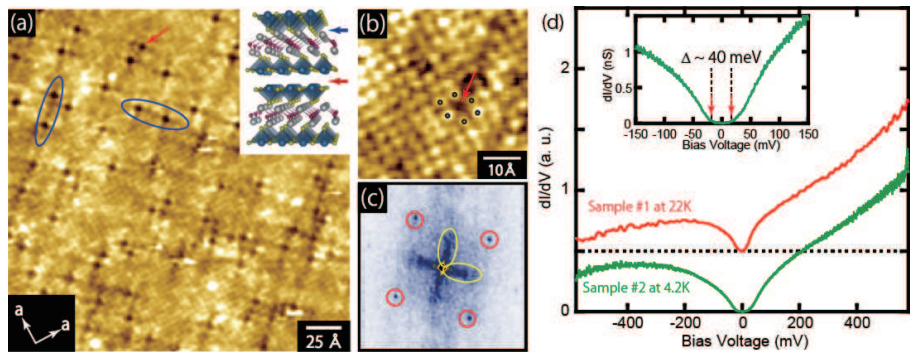}
\end{center}
\caption{
(Color online)
(a) A typical STM image on a 240 $\times$ 240 \AA$^{2}$ field of view (FOV) taken at $V_{\mathrm{s}}$ = +500 mV and $I_{set}$ = 500 pA.
The inset illustrates a schematic figure of the crystal structure, where red, gray, yellow, and blue spheres represent Nd, O, S, and Bi atoms, respectively.
(b) A magnified STM image on a 45 $\times$ 45 \AA$^{2}$ FOV.
(c) A Fourier transform image of (a). Red circles indicate the spots corresponding to the atomic array. Yellow ellipsoids mark the tails corresponding to the dark streaks running along the diagonal (110) direction of the unit cell.
(d) Typical tunneling spectra taken at $V_{\mathrm{s}}$ = +500 mV and $I_{\mathrm{s}}$ = 1.0 nA. Red (upper) and green (lower) lines represent the spectra taken on sample \# 1 at 22 K and sample \# 2 at 4.2 K. The red spectrum is shifted by 0.5 for clarity. The dashed line indicates the zero conductance for the red (shifted) spectrum. The inset shows the spectrum with high energy resolution.}
\label{Fig_1}
\end{figure*}

However, the results obtained using these surface sensitive probes occasionally do not reflect intrinsic bulk nature but reflect the nature peculiar to the surface which is sometimes a stage of a novel phenomenon, when the structure and electronic nature on the surface are different from those in the bulk. Hence, it is necessary to uncover what type of surface structure and electronic states are realized on an exposed surface, as a first step in investigating the electronic nature of new compounds.

In this study, we performed STM and STS experiments on a cleaved surface of NdO$_{0.7}$F$_{0.3}$BiS$_{2}$ ($T_{\mathrm{c}}\sim$5.1 K), which is a member of the BiS$_2$ superconductor family.
A typical STM image on a cleaved surface shows several features: (i) a square lattice with a period of $\sim$3.9 \AA\ corresponding to the in-plane lattice parameter, (ii) atomic defects, and (iii) streaked depressions that are centered on each atomic defect.
The tunneling spectroscopy captures several unexpected phenomena. Its first feature is a large spectroscopic gap ($\sim$40 mV) without a coherence peak, which survives well above $T_{\mathrm{c}}$. It is considered that the observed gap is inherent nature in the cleaved surface rather than in the bulk, since the gap opening at approximately $E_{\mathrm{F}}$ is not consistent with the metallic behavior in the bulk electric resistivity measurements. The second feature is a square lattice characterized by an alternating arrangement of two types of nanocluster. In one cluster, one-dimensional stripe electronic structures run along one Bi-Bi direction, whereas, in the other cluster, the stripes run along the other Bi-Bi direction. The third is a nanoscale electronic inhomogeneity, which might have originated from the inhomogeneous distributions of Bi defects and doped F atoms.

The single crystals used in this study were grown by a high-temperature flux method\cite{MNagao}.
The superconducting transition temperature of 5.1 K was determined by magnetization and resistivity measurements\cite{MNagao}.
The STM and STS measurements were performed in a helium-gas environment at 4.2 and 22 K using a laboratory-built STM.
A surface on which the measurements were carried out was prepared by cleaving the sample \textit{in situ}.
A bias voltage was applied to the sample in all measurements.
In this study, tunneling spectra were obtained by numerically differentiating the $I$-$V$ characteristics measured at each location.

\begin{figure*}[tb]
\begin{center}
\includegraphics[width=15cm]{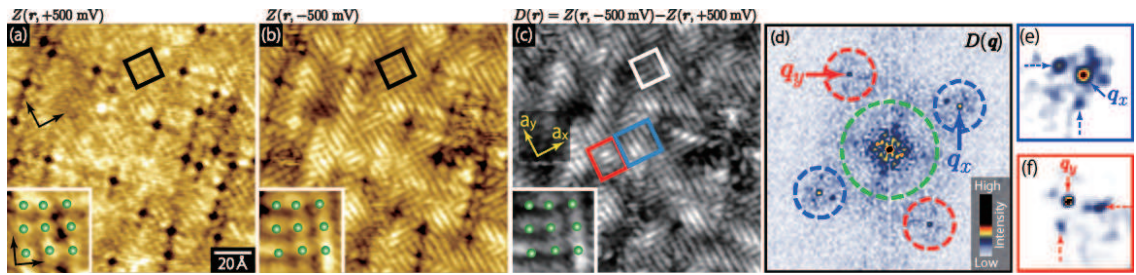}
\end{center}
\caption{
(Color online)
(a) and (b) STM images on a 130 $\times$ 130 \AA$^{2}$ field of view (FOV) taken at $V_{\mathrm{s}}$ = +500 and -500 mV.
(c) Difference between (a) and (b) [$D(\boldmath{r})$-map]; $D(\boldmath{r}) = Z(\boldmath{r}, -500\ \mathrm{mV}) - Z(\boldmath{r}, +500\ \mathrm{mV})$.
Red and blue boxes indicate the representatives of the nanodomains where the stripe structures runs along $a_{x}$- and $a_{y}$-directions, respectively.
Insets of (a), (b), and (c) are magnified images in the FOV marked by black and white boxes in (a) to (c), respectively. These magnified images are rotated by 40$^{\circ}$ in a clockwise fashion. Green spheres represent Bi atom sites.
(d) Fourier transform image [$D(\boldmath{q})$] of (c). Red and blue dashed circles indicate the size of $\Lambda$ (cutoff frequency) in the inverse FT analyses for the stripe structures. The green dashed circle indicate the size of $\Lambda$ for an inhomogeneous structure.
(e) and (f) Magnified images of (d) at around $\boldmath{q}_{\mathrm{x}}$ and $\boldmath{q}_{\mathrm{y}}$. Here, we increase the contrasts in these magnified images for clarity. Dashed arrows in these images indicate satellite peaks.
}
\label{Fig_2}
\end{figure*}

We show typical STM images in Figs. 1(a) and 1(b) indicating a square lattice with a period of $\sim$3.9 \AA\ corresponding to the in-plane lattice constant ($a_0$). This is confirmed from four spots in the FT image marked by red circles in Fig. 1(c).
In addition to the square lattice, several dark spots can be observed.
In the STM measurements of the layered compounds, it is quite important to identify which plane appears as an exposed surface.
In this material, there are three possible atomic planes that could be exposed: (i) NdO, (ii) S(2), and (iii) BiS(1).
If the cleavage occurs between the NdO and S(2) planes, as shown by a blue arrow in the inset of Fig. 1(a), two types of atomic plane [NdO and S(2) planes] should be exposed.
On the other hand, if the cleavage occurs between two BiS(1) layers shown by a red arrow in the inset, only one type of atomic plane [BiS(1) plane] should be exposed.
Because we obtained only one type of STM image, as shown in Fig. 1(a) and the tunneling spectrum in Fig. 1(d) in this study, we conclude that the observed surface is a BiS(1) plane.
Recent calculations for the density of states (DOS)\cite{BLi} show that the Bi-$p$ and S(1)-$p$ states predominantly contribute to the DOS and that the Bi-$p$ DOS is higher than that of S(1)-$p$ near $E_{\mathrm{F}}$.
Therefore, the observed bright spots in the STM images are considered to be Bi atoms.

Given that the observed bright spots are Bi atoms, the observed dark spots pointed by the red arrow in Figs. 1(a) and 1(b) are Bi defects, since the center of the dark spot coincides with one of the four Bi sites of the Bi square lattice, as shown in Fig. 1(b).
The density of the dark spots is about 3\% per Bi atom, which is consistent with the results of the recent single-crystal X-ray diffraction measurements\cite{Miura}.
Besides the atomic defects, slightly dark streaks run along approximately diagonal directions of the unit cell.
Representative examples are marked by two blue ellipsoids in Fig. 1(a).
The direction and length of the dark streaks somewhat vary.
This variability corresponds to the four tails extending from the center ($\boldmath{q}$ = 0) of the FT image, as marked by yellow ellipsoids in Fig. 1(c).
It appears that the dark streaks are highly related to the atomic defects.
Therefore, the dark streak structures are believed to be the structural distortion relaxing the mechanical stress due to atomic defects.

\begin{figure*}[tb]
\begin{center}
\includegraphics[width=15cm]{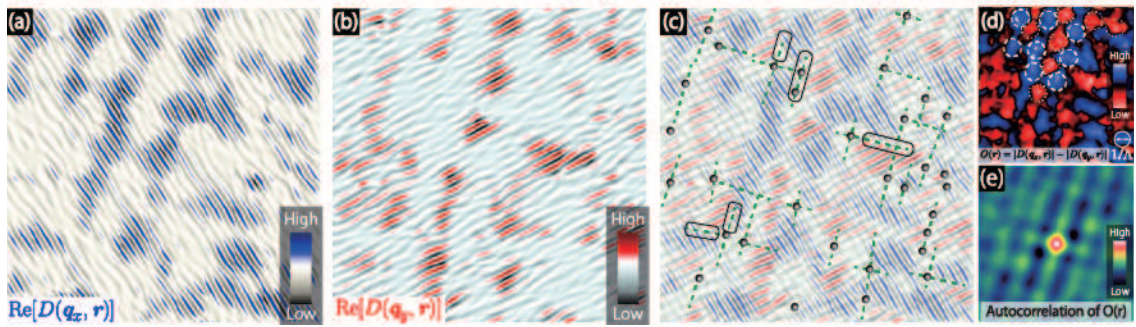}
\end{center}
\caption{
(Color online)
(a) and (b) Inverse FT images obtained by Fourier filtering out all the modulations except those surrounding $\boldmath{q}_{x}$ (a) and $\boldmath{q}_{y}$ (b): Images of the real part of $D(\boldmath{q}_{x}, \boldmath{r})$ and $D(\boldmath{q}_{y}, \boldmath{r})$, as shown in Eq. (1).
(c) Superposition of (a) and (b). Black dots and green dashed lines correspond to the atomic defects and dark streaks on the surface, respectively. Black boxes mark the dark streaks that coincide with the boundary between two nanoclusters.
(d) $O(\boldmath{r})$-map. [$O(\boldmath{r})$ = $|D(\boldmath{q}_{x}, \boldmath{r})|$ - $|D(\boldmath{q}_{y}, \boldmath{r})|$]. Dashed and dotted circles indicate a representative square lattice, which partially appears in (d). (e) Autocorrelation image of (d).
}
\label{Fig_3}
\end{figure*}

Figure 1(d) shows typical tunnelling spectra at 4.2 and 22 K, which were measured in different samples with different tips.
Both spectra indicate (i) a large energy asymmetry characterized by the higher conductance in positive energy than in negative energy, and (ii) a spectroscopic gap characterized by a depression of the spectral weight at approximately $E_{\mathrm{F}}$\cite{Tunnel_spe}.
Surprisingly, the conductance is approximately zero between $\pm$20 meV at both 4.2 and 22 K, as shown in the inset of Fig. 1(d).
This means that there is no DOS at around $E_{\mathrm{F}}$.
The observed gap is not considered a superconducting gap, because it survives well above $T_{\mathrm{c}}$ and does not have a clear coherence peak.
These spectroscopic features are inconsistent with the bulk properties showing a clear superconducting transition and a metallic behavior of the resistivity just above $T_{\mathrm{c}}$\cite{MNagao}.
Therefore, it is preferable to consider that the observed spectroscopic features do not reflect the bulk property but reflect the property unique to the surface.

To explore the spatial variation in the electronic structure, we compare two STM images taken at $V_{\mathrm{s}}$ = +500 and -500 mV on the same FOV, as shown in Figs. 2(a) [$Z$($\boldmath{r}$, +500 mV)] and 2(b) [$Z$($\boldmath{r}$, -500 mV)], respectively.
Apparently, there are two features that are more pronounced in the image at $V_{\mathrm{s}}$ = -500 mV than at +500 mV: (i) Fine stripe structures along the two $a$-axes and (ii) inhomogeneous contrasts on the background.
We take the difference between these images [$D$($\boldmath{r}$) = $Z$($\boldmath{r}$, -500 mV) - $Z$($\boldmath{r}$, +500 mV)], as shown in Fig. 2(c).
This process cancels the geometric information and extracts the electronic structures.
In this case, the $D(\boldmath{r})$-map indicates an energy asymmetry of the energy-integrated LDOS from 0 to $\pm$ 500 meV.
The two features in the $Z$($\boldmath{r}$, -500 mV) map are emphasized in the $D(\boldmath{r})$-map.
Therefore, these two features are derived from the electronic spatial variation.
The FT image of the $D(\boldmath{r})$-map in Fig. 2(d) ensures the existence of these two electronic features.
The four spots marked by red and blue dashed circles correspond to the fine stripe structures,
whereas the broad structure around $\boldmath{q}$ = 0 marked by a green dashed circle corresponds to the inhomogeneous structure.

First, we focus on the fine stripe structures being elongated along one Bi-Bi direction. As described above, the stripe structures can be classified into two types. One modulates along the $a_{\mathrm{x}}$-direction marked by a blue box in Fig. 2(c) ($x$-modulation), whereas the other modulates along the $a_{\mathrm{y}}$-direction shown by a red box ($y$-modulation).
These modulations form two types of nanoscale cluster: In one cluster, the $x$-modulation predominates, whereas, in the other, the $y$-modulation predominates.
The FT image of the $D(\boldmath{r})$-map corroborates the presence of two modulations with a period of $a_0$, as shown in Fig. 2(d). Here, we name the $\boldmath{q}$ vectors corresponding to these two modulations as $q_{\mathrm{x}}$ and $q_{\mathrm{y}}$, respectively.
Moreover, it appears that the two types of cluster tend to be aligned alternatingly and periodically in Fig. 2(c).
In the FT image, in addition to $q_{\mathrm{x}}$ and $q_{\mathrm{y}}$, two satellite peaks can be seen at around $q_{\mathrm{x}}$ and $q_{\mathrm{y}}$, as shown by dashed arrows in Figs. 2(e) and 2(f).
These satellites indicate that the amplitudes of the observed $x$- and $y$-modulations periodically and two-dimensionally vary in real space: these satellites are indicative of the alternating arrangement of the observed two types of cluster.
The period of the amplitude modulations can be estimated from $\delta \boldmath{q}$ = $\boldmath{q}_{\mathrm{x\ or\ y}}$ - $\boldmath{q}_{\mathrm{st}}$, where $\boldmath{q}_{\mathrm{st}}$ is the position of the satellite peaks.
The value of $\delta \boldmath{q}$ is approximately 0.185$\times$2$\pi$/$a_{0}$ [0.245$\times$2$\pi$/($\sqrt{2}a_{0}$)].
This means that the period of the amplitude modulation is approximately 5.4$a_{0}$ (4.1$\tilde{a}_{0}$, $\tilde{a}_{0}$ is the length between Bi atoms along the diagonal direction of the unit cell).

To visualize the corrugation of these clusters more clearly, we extract the components associated with the $x$- and $y$-modulations from the $D(\boldmath{r})$-map using the inverse Fourier transformation as follows:
\begin{eqnarray}
D(\boldmath{q}_{\mu}, \boldmath{r}) = \frac{2\pi}{N}\sum_{\boldmath{q}}D(\boldmath{q})\exp(-i\boldmath{q}\cdot\boldmath{r}) \nonumber \\
\exp\left[-(\boldmath{q}_\mu-\boldmath{q})^2/\Lambda ^2\right],\nonumber \\
\mu = x\ \mathrm{or}\ y,
\end{eqnarray}
where $\Lambda$ is the cutoff wave number.
Here, we set $\Lambda$ = 0.294 $\times$ 2$\pi$/$a_{0}$ (1/$\Lambda$ = 13.5 \AA) corresponding to the red and blue dashed circles in Fig. 2(d), as $\Lambda$ becomes longer than the $|\delta \boldmath{q}|$.
Figures 3(a) and 3(b) are the real parts of $D(\boldmath{q}_{x}, \boldmath{r})$ and $D(\boldmath{q}_{y}, \boldmath{r})$, respectively.
A region with a large amplitude of the $x$-modulation shows a small amplitude of the $y$-modulation and vice versa.
This is also confirmed more clearly in Fig. 3(c) in which Fig. 3(a) is superimposed on Fig. 3(b).
We further map out a quantity $O(\boldmath{r})$ = $|D(\boldmath{q}_{x}, \boldmath{r})|$ - $|D(\boldmath{q}_{y}, \boldmath{r})|$, as shown in Fig. 3(d). [$|D(\boldmath{q}_{x}, \boldmath{r})|$ ($|D(\boldmath{q}_{y}, \boldmath{r})|$) means the spatial variation in the amplitude of the $x$-modulation ($y$-modulation)]. In this map, the nearly 4$\tilde{a}_{0}$ square lattice of the nanoclusters can be partially observed as expected from the satellite peaks in the FT image of the $D(\boldmath{r})$-map, as marked by white dashed circles in Fig. 3(d). 
Figure 3(e) shows the autocorrelation image of the $O(\boldmath{r})$-map.
In this image, the nearly 4$\tilde{a}_{0}$ square lattice is clearly observed.
Thus, our results suggest that the observed nanoclusters tend to form a `checkerboard'-like 4$\tilde{a}_{0}$ square lattice.
The formation of the lattice implies that some interactions exist between the nanoclusters.

Figure 3(c) shows the spatial correlation among the atomic defects, the dark streak structures, and the observed nanoclusters. There is a weak correlation between the atomic defects and the stripe structures in these nanoclusters. The stripe structure tends to be suppressed around the atomic defects. The positions of the dark streak structures sometimes coincide with the boundary between two nanoclusters, as marked by the black boxes in Fig. 3(c).
This means that the dark streak structure tends to pin the boundaries of these clusters.
Thus, these local perturbations seem to distort the nearly 4$\tilde{a}_{0}$ square lattice.

The observed electronic features including a large spectroscopic gap and `checkerboard stripe' electronic structure have not been anticipated in previous investigations of the bulk properties.
It is considered that the observed spectroscopic gap seems to be inherent in the cleaved surface, because of the inconsistency between the gap opening and the metallic behavior in bulk transport measurements.
If the observed checkerboard stripe structure is related to the large spectroscopic gap, the checkerboard stripe structure is the inherent phenomenon in the cleaved surface.
However, it is also possible to think that the checkerboard stripe coexists with the bulk superconductivity, if it resides not only on the surface but also deep inside the sample.
Even though we cannot clarify whether the checkerboard stripe exists only on the surface or in the bulk at present, our findings will be crucial for understanding the electronic nature of this material, if the checkerboard stripe coexists with the superconductivity.

In discussing the origin of the checkerboard stripe, it is important to consider what is a driving force to generate the stripe structure which breaks $C_{4}$-symmetry.
One possible candidate is the softening of the phonons.
Recent calculations of the phonon spectra of LaO$_{1-x}$F$_{x}$BiS$_2$ predicted that it is possible for S(1) atoms to move towards Bi atoms along the $a$-axis (or $b$-axis) owing to the unstable phonon at around $\Gamma$ under the proper conditions\cite{TYildirim}.
If this distortion occurs, the unit cell no longer maintains the tetragonal symmetry ($C_4$-symmetry) and takes an orthogonal symmetry instead.
Such a broken symmetry of the crystal structure lifts the degeneracy of the Bi6-$p_{x}$ and Bi6-$p_{y}$  bands, which seem to lie near $E_\mathrm{F}$. This nonequivalent nature between the Bi6-$p_{x}$ and Bi6-$p_{y}$ bands seems to induce the observed unidirectional electronic structure (stripe structure).
However, even if we adopt this idea, several questions still remain.
Why do the nanoclusters form the nearly 4$\tilde{a}_{0}$ square lattice?
Is the condition of the exposed surface obtained by cleaving the sample is appropriate for inducing such unstable phonons?
Thus, the origin of the observed `checkerboard stripe' electronic structure remains an open question.

\begin{figure}[th]
\begin{center}
\includegraphics[width=7cm]{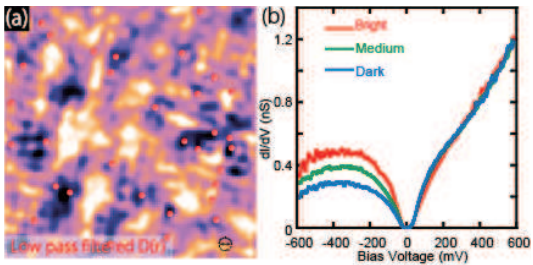}
\end{center}
\caption{
(Color online)
(a) Low-pass-filtered image of Fig. 1(c). In this low pass filtering, we set the cutoff frequency to be $|\boldmath{q}|$ = 0.58 $\times$ 2$\pi$/$a_{0}$ corresponding to the real-space cutoff radius of 6.7 \AA\ indicated by a black circle on the bottom-right corner. Red dots show the positions of the atomic defects.
(b) Tunneling spectra at several locations. Red, green, and blue lines indicate the spectrum taken in the bright, medium contrast, dark regions in (a).
}
\label{Fig_4}
\end{figure}

Next, we focus on inhomogeneous structures.
Figure 4(a) shows a low-pass-filtered image of the $D$($\boldmath{r}$)-map, where the cutoff wave number is $q_{\mathrm{cut}}$ = 0.58 $\times$ 2$\pi$/$a_{0}$ (corresponding to a cutoff radius $r_{\mathrm{cut}}$ = 6.7 \AA).
This filtering subtracts the fine stripe structures and extracts the electronic inhomogeneity.
This filtered image contains inhomogeneous spatial variations of the energy integrated LDOS in both $Z$($\boldmath{r}$, -500 mV) and $Z$($\boldmath{r}$, +500 mV).
In this case, however, the contribution of $Z$($\boldmath{r}$, -500 mV) appears to be larger than that of $Z$($\boldmath{r}$, +500 mV), because there is a larger contrast regarding the inhomogeneous spatial variation in $Z$($\boldmath{r}$, -500 mV) than in $Z$($\boldmath{r}$, +500 mV), as shown in Figs. 2(a) and 2(b).
Thus, the bright region in the filtered $D$($\boldmath{r}$)-map seems to correspond to the region with a higher DOS in negative energy.
To further explore how the shape of the tunneling spectrum changes along the observed inhomogeneity, we take tunneling spectra at several locations with the different intensities of $D$($\boldmath{r}$), as shown in Fig. 4(b).
Here, we set the set bias voltage $V_{\mathrm{s}}$ to be +500 mV, because the spatial variation in the energy-integrated DOS is smaller in $Z$($\boldmath{r}$, +500 mV) than in $Z$($\boldmath{r}$, -500 mV).
This set-up condition appears to alleviate a set point effect on the tunneling spectrum.
An apparent change in the spectral weight lies below -100 meV.
Thus, the LDOS below -100 meV is closely related to the source of the inhomogeneity.

There are several possible sources of the electronic inhomogeneity. The first candidate is the inhomogeneous distribution of the atomic defects. As seen in Fig. 4(a), the defects prefer to sit in the dark region with the lower DOS in negative energy. This slight correlation implies that the atomic defects are related to the electronic inhomogeneity. These atomic defects would induce the structural distortions around them. The distortions induced by the atomic defects might provide a local modification of the hopping integrals, which alters the electronic states locally. Another candidate is the inhomogeneous distribution of the dopant F atoms. Usually, the dopant atoms distribute in a random fashion. The random distribution of the dopant F atoms might provide the inhomogeneous distribution of the carrier density, which also seems to cause the electronic inhomogeneity.

In summary, we report on our STM and STS experiments on a cleaved surface of NdO$_{0.7}$F$_{0.3}$BiS$_{2}$ with $T_{\mathrm{c}}$ = 5.1 K.
The STM images contain a square lattice with a period of $\sim$3.9 \AA\ corresponding to the lattice parameter along the $a$-axis and the several atomic defects.
The tunneling spectra indicate a large spectroscopic gap ($\sim$40 mV), which is not the superconducting gap.
It is conceivable that the observed spectroscopic gap at $E_{\mathrm{F}}$ is inherent in the cleaved surface, because the metallic behavior in the bulk transport measurements is not consistent with the gap opening.
In addition to this spectroscopic gap, we find two unexpected electronic structures: (i) `checkerboard stripe' electronic structure and (ii) a nanoscale electronic inhomogeneity.
The former is characterized by an alternating arrangement of two types of nanocluster. In one cluster, one-dimensional stripe structures run along one of the crystal $a$-axes with a period of $a_{0}$. In the other cluster, the stripes run along the other $a$-axis.
Furthermore, these clusters form a square lattice with a period of nearly 4$\tilde{a}_{0}$ along the diagonal directions of the unit cell.
The latter seems to be related to the inhomogeneous distributions of the atomic defects and of the dopant F atoms.
Our results in this study will play an important role in considering the origin of the superconductivity in this system, if the observed electronic spatial structures coexist with the bulk superconductivity.


\begin{thebibliography}{99}
\bibitem{Tunnel_spe}
In this study, we performed spectroscopy experiments on four samples and obtained the tunneling spectra in various areas (about ten areas) of each sample. All the tunneling spectra obtained in these experiments show the same spectroscopic features characterized by a large energy asymmetry and the existence of a large spectroscopic gap, as shown in Fig. 1(d).
\end{thebibliography}
\end{document}